\documentclass{moriond}

\def\be{\begin{equation}}
	\def\ee{\end{equation}}
\def\bea{\begin{eqnarray}}
	\def\eea{\end{eqnarray}}

\usepackage{amsfonts}

\usepackage{bm}
\usepackage{hyperref}
\usepackage{mathrsfs}
\usepackage{graphicx}
\usepackage{empheq}
\usepackage{ulem}

\usepackage[usenames]{color}
\usepackage{ytableau}

\DeclareSymbolFontAlphabet{\mathrsfs}{rsfs}
\DeclareMathAlphabet{\mathcal}{OMS}{cmsy}{m}{n}

\newcommand{\dd}{\mathrm{d}}

\newcommand{\Qcal}{{\cal Q}}

\newcommand{\bse}{\begin{subequations}}
	\newcommand{\ese}{\end{subequations}}

\newcommand{\AEST}{\text{AeST}}
   
\newcommand{\grad}{\ensuremath{\vec{\nabla}}}

\newcommand{\wt}{\ensuremath{\tilde{w}}}

\newcommand{\rC}{\ensuremath{r_C}}
\newcommand{\rM}{\ensuremath{r_M}}

\newcommand{\PsiM}{\Psi}
\newcommand{\cs}{ c_s}
\newcommand{\cvis}{ c_{\rm vis}}

\newcommand{\Qcalb}{\overline{{\cal Q}}}

\newcommand{\Ocal}{{\cal O}}
\newcommand{\Jcal}{\ensuremath{{\cal J}}}

\newcommand{\Ycal}{{\cal Y}}

\newcommand{\lambdad}{\lambda_{D}}
\newcommand{\taub}{\overline{\tau}} 
 
\newcommand{\Kcal}{{\cal K}} 
 
\newcommand{\Kcalb}{\overline{{\cal K}} }
\newcommand{\rhob}{\overline{\rho} }
\newcommand{\Pb}{\overline{P} }
\newcommand{\rhoM}{{\rho_m}}

\newcommand{\cad}{c_{\rm ad}}

\definecolor{orange}{rgb}{1,0.5,0}
\definecolor{darkorange}{rgb}{0.69,0.33,0.13}
\definecolor{fidcol}{rgb}{0.7,0,0}

\newcommand{\Photo}{}

\begin{document}
\vspace*{4cm}
\title{Khronon-Tensor theory reproducing MOND and the cosmological model}
	
\author{Luc Blanchet$^{a}$, Constantinos Skordis$^{b,c}$}
	
\address{$^{a)}$ $\mathcal{G}\mathbb{R}\varepsilon{\mathbb{C}}\mathcal{O}$, 
		Institut d'Astrophysique de Paris, 
		98\textsuperscript{bis} boulevard Arago, 75014 Paris, France\\
		$^{b)}$ CEICO, Institute of Physics of the Czech Academy of Sciences, 
		Prague, Czech Republic\\
	    $^{c)}$ Department of Physics, University of Oxford, 
	    Oxford OX1 3RH, UK}
	
\maketitle\abstracts{We propose an alternative scalar-tensor theory based on the Khronon scalar field labeling a family of space-like three-dimensional hypersurfaces. This theory leads to modified Newtonian dynamics (MOND) at galactic scales for stationary systems, recovers GR plus a cosmological constant in the strong field regime, and is in agreement with the standard cosmological model and the observed cosmic microwave background anisotropies.}

\section{Introduction}

The success of the ``MOdified Newtonian Dynamics'' (MOND) empirical formula~\cite{Milg1} at matching many measurements of dark matter at galactic scales, remains unexplained. Extensions of general relativity (GR), aiming at reproducing the MOND formula in galaxies without the need of dark matter, were proposed. The classic example is the Tensor-Vector-Scalar (TeVeS) theory~\cite{Bek04}, which however falls short at explaining the observed cosmology~\cite{Sk06}. More recently the Aether-Scalar-Tensor ($\AEST$) theory~\cite{SZ21} has been the first GR extension able to reproduce the MOND formula in galaxies and simultaneously being in agreement with the standard cosmological model. The previous theories are tensor-vector-scalar theories, and we present here (based on Ref.~\cite{BS24}, after previous attempts in Refs.~\cite{BM11} and~\cite{Sand11}) a simpler theory, without need of an independent vector field, dubbed Khronon-Tensor (KT) theory. The Khronon field is associated to a foliation of space-time by three dimensional spatial hypersurfaces, implying a violation of the local Lorentz invariance. The KT theory was inspired by the possibility of breaking the local Lorentz invariance at high energy to help the problem of quantization of gravity~\cite{Horava2009}, but considered here in a completely different context, that of the low energy, weak acceleration regime of MOND, below the MOND acceleration scale $a_0\simeq 1.2\,10^{-10}\,\mathrm{m}/\mathrm{s}^2$.
	
\section{The Khronon-Tensor theory}\label{Sec:Action}
	
The covariant theory's action is
\begin{align}\label{K_action}
	S =  \frac{c^3}{16\pi G}\int \dd^4x \,\sqrt{-g} \,\Bigl[ R - 2 \Jcal(\Ycal) + 2 \Kcal(\Qcal) \Bigr] 
		+ S_m\left[\PsiM,g\right]\,,
\end{align}
where $R$ is the Ricci scalar, and the standard matter fields $\PsiM$ (baryons) are universally coupled to the metric $g_{\mu\nu}$. The theory extends GR with the addition of two scalar functions.
%

First, the function $\Jcal(\Ycal)$ is postulated in order to recover the MOND formula in the low acceleration regime.
It is defined from the space-time foliation by three-dimensional hypersurfaces, through the orthonormal unit-timelike vector field $n_\mu$ of the foliation,
\begin{align}\label{nmu_def}
	n_\mu = - \frac{c}{\Qcal} \nabla_\mu \tau\,,\quad\text{with}\quad\Qcal = c\sqrt{- g^{\mu\nu} \nabla_\mu \tau \nabla_\nu \tau}\,,
\end{align}
where $\tau$ is the Khronon scalar field, and the acceleration of the congruence of unit vector $n_\mu$ reads $A_\mu =  c^2 n^\nu \nabla_\nu n_\mu$, or equivalently
\begin{align}\label{Amu_def}
	A_\mu =  - c^2 q_{\mu}^{\;\;\nu}  \,\nabla_\nu \ln \Qcal \,,\quad\text{with}\quad q_{\mu}^{\;\;\nu} = \delta_\mu^{\nu}  + n_\mu n^\nu\,.
\end{align}
Posing then $\Ycal = A_\mu A^\mu/c^4$, we assume that the function $\Jcal(\Ycal)$ vanishes in the high acceleration limit, 
while it takes a following specific form in the low acceleration limit:
%
\begin{subequations}\label{Jvalue}
\begin{align}
	&\Jcal = 0 &\text{when}\quad\Ycal \gg \frac{a_0^2}{c^4}\,,\\
	&\Jcal = \Lambda - \Ycal + \frac{2c^2}{3 a_0}\,\Ycal^{3/2}+ \Ocal\left(\Ycal^2\right)&\text{when}\quad\Ycal \ll \frac{a_0^2}{c^4}\,.\label{highaccel}
\end{align}
\end{subequations}
Note that the low acceleration regime of MOND is also the regime of linear cosmological perturbations, since the acceleration $A_\mu$ is a first order quantity in cosmological perturbations. Thus, $\Ycal$ is second order in cosmology and therefore, the constant $\Lambda$ in~\eqref{highaccel} is the measured cosmological constant.

Second, the function $\Kcal(\Qcal)$, where $\Qcal$ is given in~\eqref{nmu_def}, serves as a kinetic term for the Khronon field and will ensure the consistency with cosmological observations. Following~\cite{SZ21}, our main assumption is that this function should admit a Taylor expansion around the value $\Qcal=1$. For definiteness we can adopt a simple quadratic potential,
\begin{align}
	\Kcal(\Qcal) = \mu^2\left(\Qcal -1\right)^2 \,,
	\label{K_def}
\end{align}
where $\mu$ is a constant with dimension of an inverse length. The term~\eqref{K_def} plays a crucial role in the success of $\AEST$ theory in fitting cosmological observations~\cite{SZ21}. We shall also consider a full non-perturbative expansion inspired by the Dirac-Born-Infeld (DBI) function,
\begin{align}\label{DBIfunction}
	\Kcal_\text{DBI}(\Qcal) = \frac{2\mu^2}{\lambdad} \left[1 - \sqrt{1 - \lambdad\left(\Qcal-1\right)^2 }  \right]\,,
\end{align}
where $\lambdad$ is a second parameter, and show that it actually fits better all observations.
%
%
	 
\section{Post-Newtonian limit}
	
In the PN or slow motion approximation, we introduce two \textit{a priori} different scalar potentials $\phi$ and $\psi$, as well as the ``gravitomagnetic'' vector potential $\zeta_i$, and make the usual PN ansatz on the metric components generated by an isolated system,
\begin{subequations}\label{PNmetric}
\begin{align}
		g_{00} &= -1 - \frac{2\phi}{c^2} + \Ocal(  c^{-4} )\,, \\
		g_{0i} &= \frac{4}{c^3}\,\zeta_i +\Ocal( c^{-5} )\,, \label{PNNi}\\
		g_{ij} &= \gamma_{ij}\Bigl(1 - \frac{2\psi}{c^2}\Bigr) + \Ocal( c^{-4} ) \,,
\end{align}
\end{subequations}
where $\gamma_{ij}$ is a Euclidean metric in general coordinates, and with usual notation for the PN remainders $\Ocal( c^{-n})$. The PN ansatz~\eqref{PNmetric} is standard in GR and follows from the leading PN order of the stress-energy tensor $T^{\mu\nu}$ for ordinary matter fields. However, for the present KT theory we must also take into account the Khronon field equation, and ensure that the stress-energy tensor of the Khronon field admits the same leading PN behaviour as for the ordinary matter. This will be satisfied if the expansion of the Khronon field is of the type~\cite{Flanagan23}
\begin{align}\label{tausigma}
	\tau = t + \frac{\sigma(\mathbf{x},t)}{c^2} + \Ocal( c^{-4} )\,,
\end{align}
where $\sigma$ is the leading order perturbation. Plugging~\eqref{PNmetric} and~\eqref{tausigma} into the Einstein field equations, using the facts that $\Jcal$ and $\Kcal$ are small PN quantities $\Ocal(c^{-4})$ (see Ref.~\cite{BS24} for details), we obtain, as the consequence of the $ij$ components of the field equations, 
\begin{align}\label{phipsi}
	\phi = \psi + \Ocal( c^{-2})\,.
\end{align}
The equality of the two potentials $\phi$ and $\psi$ in the PN limit is very important for the viability of the theory, as it implies that the light deflection and the gravitational lensing are given by the same formula as in GR. That is, for any baryonic distribution of matter for which the forces $\grad\phi$ are consistent with observations as if dark matter was present, the same matter distribution will also lead to the correct gravitational lensing signal, again as if dark matter was present. 
%
Next we obtain the $00$ and $0i$ components of the Einstein field equations as
\begin{subequations}
	\label{EE}
	\begin{align}
		\label{EEa} \Delta\phi &=  4\pi G \bigl(\rhoM + \rho_\tau \bigr) + \Ocal( c^{-2} )\,,
		\\[0.2cm]
		\Delta\zeta^i - \grad^i \grad_j \zeta^j  - \grad^i\dot{\phi} &= 4\pi G \left(\rhoM v_m^i + \rho_\tau v_\tau^i\right) + \Ocal( c^{-2})\,,
	\end{align}
\end{subequations}
where we have defined the Khronon ``mass density'' and ``velocity'' field (in addition to the matter contributions $\rhoM$ and $v_m^i$)
\begin{equation}
		\label{khrononrhov}	
		\rho_\tau = -\frac{1}{4\pi G}\left[ \grad \cdot \left( \Jcal_\Ycal \grad \Xi\right) + \mu^2 \Xi\right] \,,\quad\text{and}\quad
		v_\tau^i = - \grad^i\sigma\,.
\end{equation}
We pose $\Xi \equiv \phi - \dot{\sigma} + \frac{1}{2}|\grad\sigma|^2$. The Einstein field equations~\eqref{EE}, together with the continuity equation for ordinary matter, imply the continuity equation for the Khronon variables, \emph{i.e.}
\begin{equation}
	\label{Khreq}
	\dot{\rho}_\tau + \grad_i\left(\rho_\tau v_\tau^i\right) = \Ocal( c^{-2})\,.
\end{equation}
In turn, this equation is nothing but the PN limit of the Khronon field equation, that can be derived from the KT action~\eqref{K_action} by varying with respect to $\tau$.

\begin{figure}[!t]
	\centerline{\includegraphics[width=0.8\textwidth]{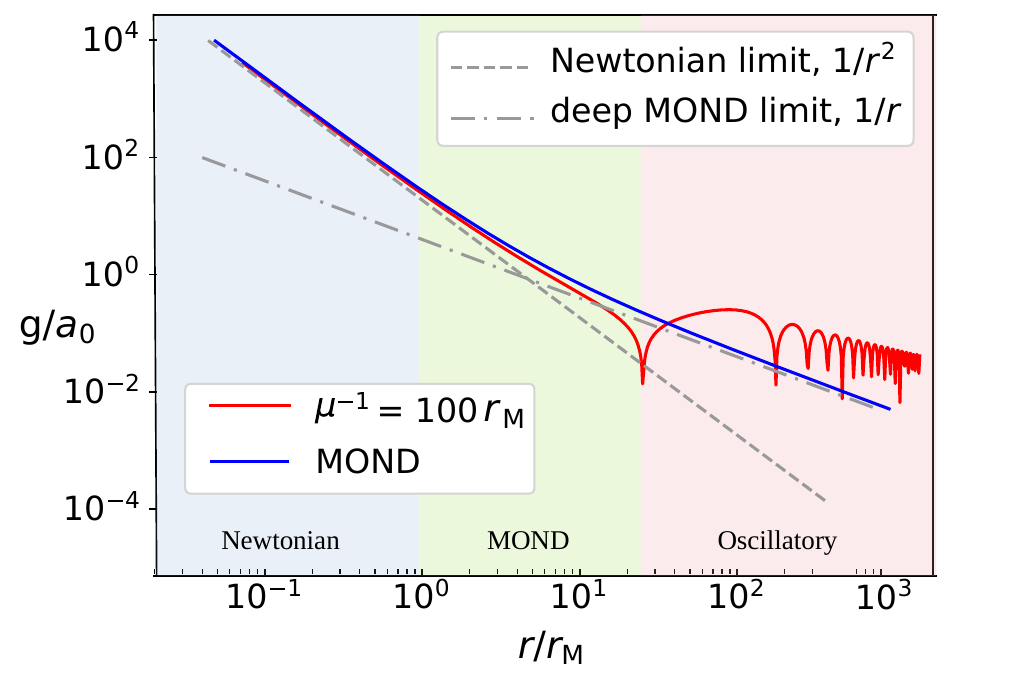}}
	\caption{Typical behaviour of the solution (red curve) of the modified MOND equation~\eqref{eqMOND}. The MOND transition occurs at the radius $\rM = \sqrt{G M/a_0}$. The transition to the oscillatory regime occurs at $\rC\sim ( \rM/\mu^2)^{1/3}$.}
	\label{Fig1}
\end{figure}

Next we restrict attention to stationary situations. Thus $\dot{\phi}=\dot{\sigma}=0$ for the fields and $\dot{\rhoM} =\dot{v}_m^i=0$ for the matter, and the continuity equation for matter reduces to the constraint $\grad_i\left(\rhoM v_m^i\right)=0$. Similarly the Khronon equation~\eqref{Khreq} reduces to $\grad\cdot(\rho_\tau \grad\sigma)=0$ which we can solve by choosing $\sigma=0$. Therefore, in the stationary case, we can actually choose the so-called unitary gauge, where the Khronon field has no perturbation and it is given by the time coordinate exactly: $\tau=t$. The acceleration then recovers the physical Newtonian acceleration, $A_i = \grad_i\phi + \Ocal( c^{-2} )$, and we find that the equation for the Newtonian potential $\phi$ becomes the modified Poisson (or modified Helmholtz) equation
\begin{equation}\label{eqMOND}
	\grad\cdot\left[  \left(1+\Jcal_\Ycal\right)\grad\phi\right] + \mu^2 \phi  = 4\pi G \rhoM + \Ocal( c^{-2} )\,.
\end{equation}
This equation takes the form of the MOND equation~\cite{BekM84} if we identify $1+\Jcal_\Ycal$ with the MOND interpolating function. By choosing Eqs.~\eqref{Jvalue} we recover both the Newtonian regime and the deep MOND regime at low accelerations. However we also find (exactly like for the $\AEST$ case) a ``mass'' term responsible for oscillations of the gravitational field at large distances from the center. The typical behaviour of the field is illustrated by the Fig.~\ref{Fig1}.

\section{The cosmological limit}

We consider a Friedmann-Lema\^{i}tre-Robertson-Walker (FLRW) background cosmology with scale factor $a$, cosmic time $t$ and a spatially homogeneous and isotropic metric of spatial curvature $\kappa$. The symmetries of the FLRW metric also require that the Khronon field be a function of time only, that is, $\tau =\taub(t)$, so that the background value of $\Qcal$ is $\Qcalb = \dot{\taub}$, with the dot denoting derivatives with respect to cosmic time $t$, and the overbar always denoting the background value.

As readily seen from the Friedmann equations, the background Khronon field contributes in the form of energy density and pressure given by
\begin{align}\label{rhobPb_Kcalb}
		\rhob_{\Kcal} = \frac{\Qcalb \,\Kcalb_{\Qcal} - \Kcalb}{8 \pi G} \,,\qquad
		\Pb_{\Kcal} = \frac{\Kcalb}{8 \pi G} \,,
\end{align}
where $\Kcal(\Qcal)$ and its derivative $\Kcal_\Qcal(\Qcal)=\dd\Kcal/\dd\Qcal$ are computed in the FLRW background. Thus, on the background, the Khronon field can be recast as a perfect fluid with time-dependent equation of state $w$ and adiabatic speed of sound $\cad^2$ given by
\begin{align}
	w \equiv \frac{\Pb_{\Kcal}}{\rhob_{\Kcal}} = \frac{\Kcalb}{ \Qcalb \,\Kcalb_{\Qcal} - \Kcalb } \,,\qquad\cad^2 \equiv \frac{\dd\Pb_{\Kcal}}{\dd\rhob_{\Kcal}} = \frac{\Kcalb_{\Qcal}}{\Qcalb\,\Kcalb_{\Qcal\Qcal}}\,.
	\label{w_Kcalb}
\end{align}
This description of the Khronon field is consistent with the energy conservation, $\dot{\rhob}_{\Kcal} + 3 H (1+w) \rhob_{\Kcal} = 0$, which is equivalent to the Khronon equation and can be integrated once to give ($I_0$ being a constant set by initial conditions)
\begin{align}
	\Kcalb_{\Qcal} =  \frac{I_0}{a^3}\,.
	\label{Kcal_I_0}
\end{align}

Specifying a function $\Kcal(\Qcal)$ completely determines the form of $\rhob_{\Kcal}$, $\Pb_{\Kcal}$, $w$ and $\cad^2$. In addition, Eq.~\eqref{Kcal_I_0} may be inverted to find a solution for $\Qcalb(a)$. When inserted into~\eqref{rhobPb_Kcalb}, this determines the exact dependence of the Khronon energy density and pressure in terms of the scale factor $a$. Our aim is to determine the conditions on $\Kcal(\Qcal)$ that may lead to approximate dust solutions for the Khronon, in accordance with observations. Let us remark that exact dust solutions, \textit{i.e.} $w=0$, are impossible as they would imply that $\Kcal = 0$. Following~\cite{Scherrer2004} and~\cite{Arkani2004} we find that approximate dust solutions exist if $\Kcal(\Qcal)$ is expandable as a Taylor series around $\Qcal=1$. This was the motivation for the specific choices made in Eqs.~\eqref{K_def} and~\eqref{DBIfunction}. For instance, with the ``quadratic'' function~\eqref{K_def} we get
\begin{equation}\label{epsilonFLRW}
	\rhob_{\Kcal} = \frac{I_0}{8\pi G a^3}\left( 1 + \frac{\wt_0}{a^3}\right)\,,\qquad w = \frac{\Qcalb - 1}{\Qcalb + 1 } = \frac{\wt_0}{\wt_0 + a^3}\,,
\end{equation}
where we pose $\wt_0  = \frac{I_0}{4\mu^2}$ and the constant $I_0$ is directly related to the Khronon energy density today (\textit{i.e.}, when $a=1$). We thus obtain approximate dust solutions in the late universe which can make the KT theory in agreement with cosmological observations, provided these dust solutions can be extended as far back as the radiation-matter equality. But we observe that the equation of state in~\eqref{epsilonFLRW} tends to $w\rightarrow 1$ (\textit{i.e.}, behaves as a stiff fluid) in the Early Universe. 

As shown in Ref.~\cite{BS24}, choosing the constant parameter $\wt_0$ in~\eqref{epsilonFLRW} to be small enough to pass the constraints from the CMB and cosmology, implies a tension with MOND, in the sense that the mass parameter $\mu$ (which rules the oscillatory behaviour of the gravitational acceleration as shown in the Fig.~\ref{Fig1}) is then too small to yield a consistent MOND solution in a typical galaxy. Fortunately, by postulating a different (non perturbative) behaviour for the function $\Kcal(\Qcal)$ we can arrange the equation of state to behave approximatively as dust not only today but also in the early universe, hence to easily pass the constraints from the CMB, while being consistent with the MOND phenomenology at galactic scales. 

\begin{figure}[!t]
	\centerline{\includegraphics[width=0.8\textwidth]{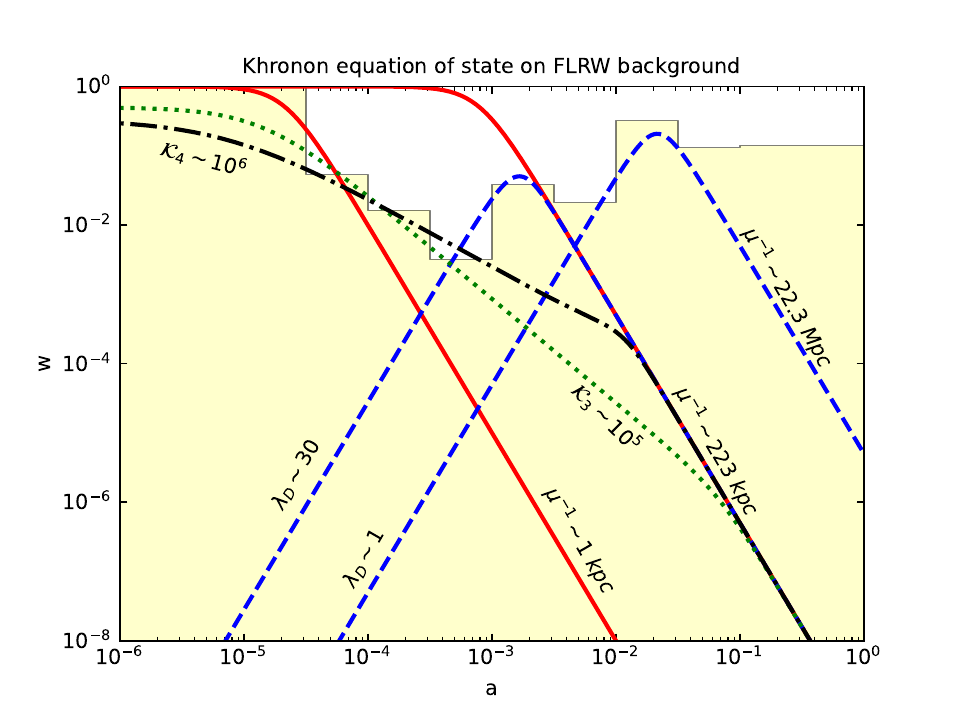}}
  \caption{The Khronon equation of state parameter $w(a)$ versus scale factor $a$ for several models. The yellow shaded region is allowed by the CMB data. The two DBI-inspired models defined by~\eqref{DBIfunction} (blue dashed curves) are compatible with both MOND and the CMB.}
	\label{Fig2}
\end{figure}

This is the case of the DBI-inspired function~\eqref{DBIfunction} which passes both constraints from the CMB and MOND for reasonable values of parameters, as shown in Fig.~\ref{Fig2}.
Finally, considering a linear perturbations on the FLRW background, we have shown that the Khronon equations can be recast into the framework of the generalized dark matter (GDM) model~\cite{Hu98}\,~\cite{KST16}. This model is defined only on the FLRW background and linearized perturbation level, and is determined by three parametric functions: the time-dependent equation of state $w(t)$, and, in the Fourier domain, the sound speed $\cs^2(t,k)$ and viscosity $\cvis^2(t,k)$ --- the last two parameters appearing at the linearized perturbed level. To conclude, the Khronon-tensor theory has the potential of reproducing in a natural way all observations both in cosmology and for galaxies.
	
\section*{References}
\bibliography{khronon_moriond}	
	
\end{document}